\documentclass[a4paper,12pt]{article}
\pdfoutput=1
\usepackage{a4wide}
\usepackage[latin1]{inputenc}
\usepackage{dsfont}
\usepackage{amsfonts}
\usepackage{amsmath}
\usepackage{amssymb}
\usepackage{amsthm}
\usepackage{graphicx}
\usepackage[numbers, sort]{natbib}
\allowdisplaybreaks[1]
\usepackage{color}
\usepackage{ulem}
\usepackage{multirow}

\usepackage[small,bf]{caption}
\newcommand{\be}{\begin{equation}}
\newcommand{\ee}{\end{equation}}
\newcommand{\beq}{\begin{equation}}
\newcommand{\eeq}{\end{equation}}
\newcommand{\bea}{\begin{eqnarray}}
\newcommand{\eea}{\end{eqnarray}}

\usepackage{commath}
\usepackage{mathcomp}
\usepackage{colortbl}
\usepackage{bbold}
\usepackage{commath}

\newcommand{\Hubble}{ \mathcal{H} }

\newcommand{\kahler}{K\"{a}hler }

\usepackage{slashed}

\usepackage[below, above]{placeins}

\begin{document}

\begin{titlepage}

\vspace*{-15mm}
\vspace*{0.7cm}

\begin{center}

{\Large {\bf Effects of the imaginary inflaton component in supergravity new inflation}}\\[8mm]

David Nolde\footnote{Email: \texttt{david.nolde@unibas.ch}}

\end{center}

\vspace*{0.20cm}

\centerline{\it
Department of Physics, University of Basel,}
\centerline{\it
Klingelbergstr.\ 82, CH-4056 Basel, Switzerland}

\vspace*{1.2cm}

\begin{abstract}
\noindent
When models of new inflation are implemented in supergravity, the inflaton is a complex and not a real scalar field. As a complex scalar field has two independent components, supergravity models of new inflation are naturally two-field models. In this paper, we use the $\delta N$ formalism to analyse how the two-field behaviour modifies the usual single-field predictions. We find that the model reduces to the single-field limit if the inflaton mass term is sufficiently small. Otherwise, the imaginary inflaton component reduces the amplitude $A_s$ and the spectral index $n_s$ of the scalar curvature perturbations. However, the perturbations remain nearly Gaussian, and the reduced bispectrum $f_{\text{NL}}$ is too small to be observed.
\end{abstract}
\end{titlepage}

\newpage

\section{Introduction}
The recent measurement of the cosmic microwave background (CMB) by the Planck collaboration \cite{Planck} has found excellent agreement with the paradigm of slow-roll, small-field inflation. In particular, the fluctuations seem to be adiabatic and Gaussian, with a slightly tilted power spectrum, confirming the generic predictions of slow-roll single-field inflation. Also, there are no signs of tensor perturbations, which is expected for small-field models, but puts tension on popular large-field models of inflation.

New inflation \cite{newInflation}, in which inflation is driven by a scalar field that slowly rolls from a hilltop towards its global minimum, is an attractive class of models which can generate CMB perturbations inside the Planck experiment's 1$\sigma$ bounds. It also allows interesting connections between inflation and particle physics. As the inflaton takes on a vacuum expectation value after inflation, it can be identified with a symmetry-breaking Higgs field which breaks e.g.\ an extended gauge symmetry \cite{BLInflation} or some family symmetry \cite{flavonInflation}. If the symmetry breaking produces any dangerous topological defects, those are produced at the beginning of inflation and are automatically diluted away by the subsequent exponential expansion. Also, the field values during new inflation stay well below the Planck scale, which makes it possible to derive predictions within an effective field theory without making strong assumptions about Planck scale physics.

However, when models of new inflation are implemented in supergravity \cite{BLInflation,flavonInflation,newInflationSUGRA1,preinflation,newInflationSUGRA2,newInflationSUGRA3}, the inflaton is a complex and not a real scalar field. As a complex scalar field has two independent components, supergravity models of new inflation are naturally two-field models. In this paper, we want to analyse how this two-field behaviour modifies the usual single-field predictions. We discuss for which model parameters the model reduces to the well-known single-field model, and how the predictions are modified otherwise.

The paper is structured as follows. First we introduce the inflaton potential, including the superpotential and \kahler potential from which the inflaton potential is derived. Afterwards, we analytically study the initial conditions and field trajectories to understand under which conditions the model reduces to a single-field model, and under which conditions we expect that the multi-field dynamics should influence the inflationary predictions. Finally, we use the $\delta N$ formalism to numerically calculate the predictions of the two-field model. We finish with a brief summary of our results.

\section{Scalar potential}
\label{sec:scalarPotential}

In this paper, we will study new inflation with the inflaton potential
\begin{align}
 V \, = \, V_0 \left\{   \Bigl| 1 - \frac{ \left( \chi + i \psi \right)^\ell }{M^2}  \Bigr|^2  -   \frac{ \beta }{ 2 } \left( \chi^2 + \psi^2 \right)   \right\} \label{eq:Vscalar}
\end{align}
for the real scalar fields $\chi$ and $\psi$, which can be thought of as components of a complex scalar field $\Phi = \frac{1}{\sqrt{2}}\left(  \chi + i \psi  \right)$.

\begin{figure}[tbp]
  \centering
$\begin{array}{cc}
\includegraphics[width=0.48\textwidth]{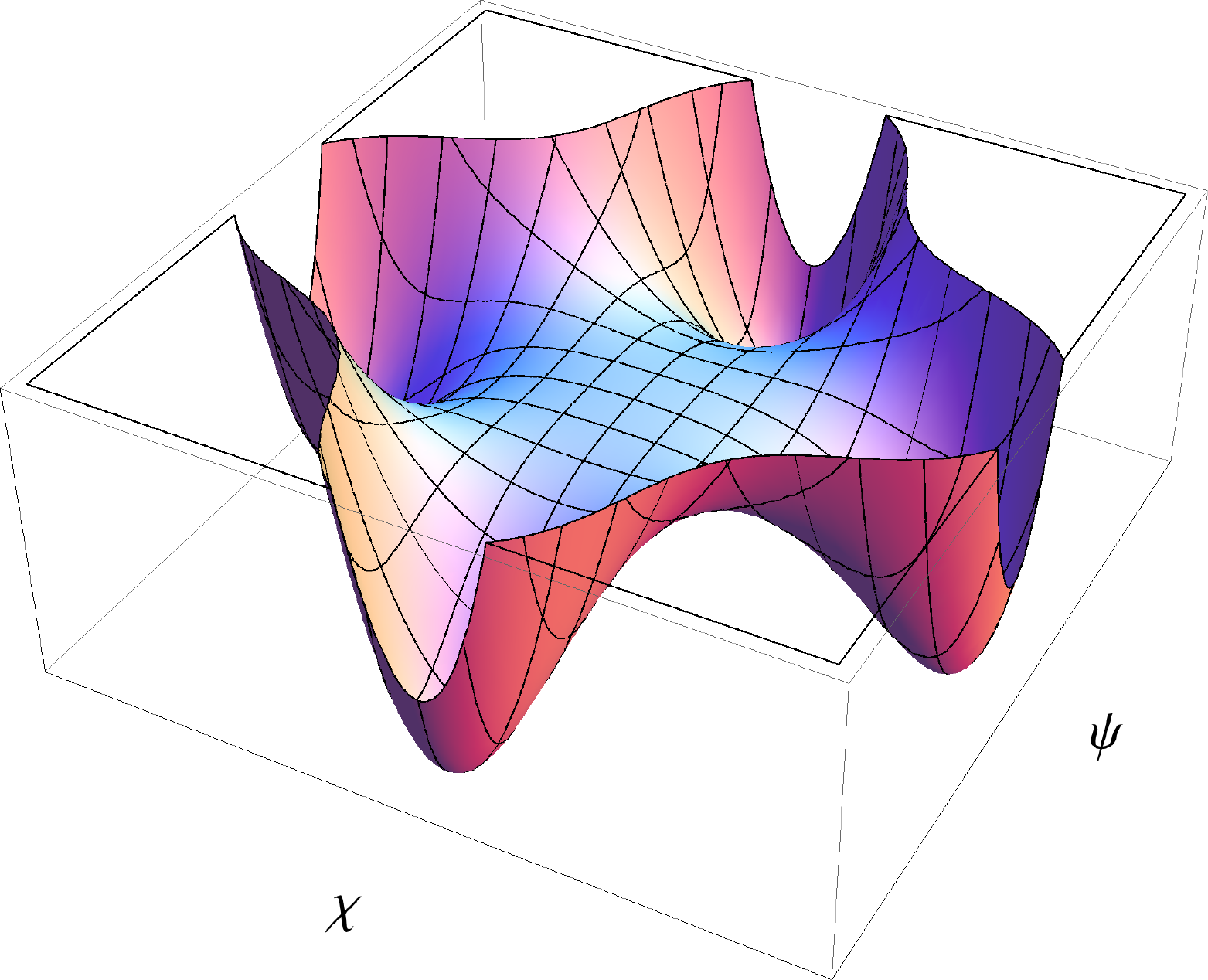} &
\includegraphics[width=0.48\textwidth]{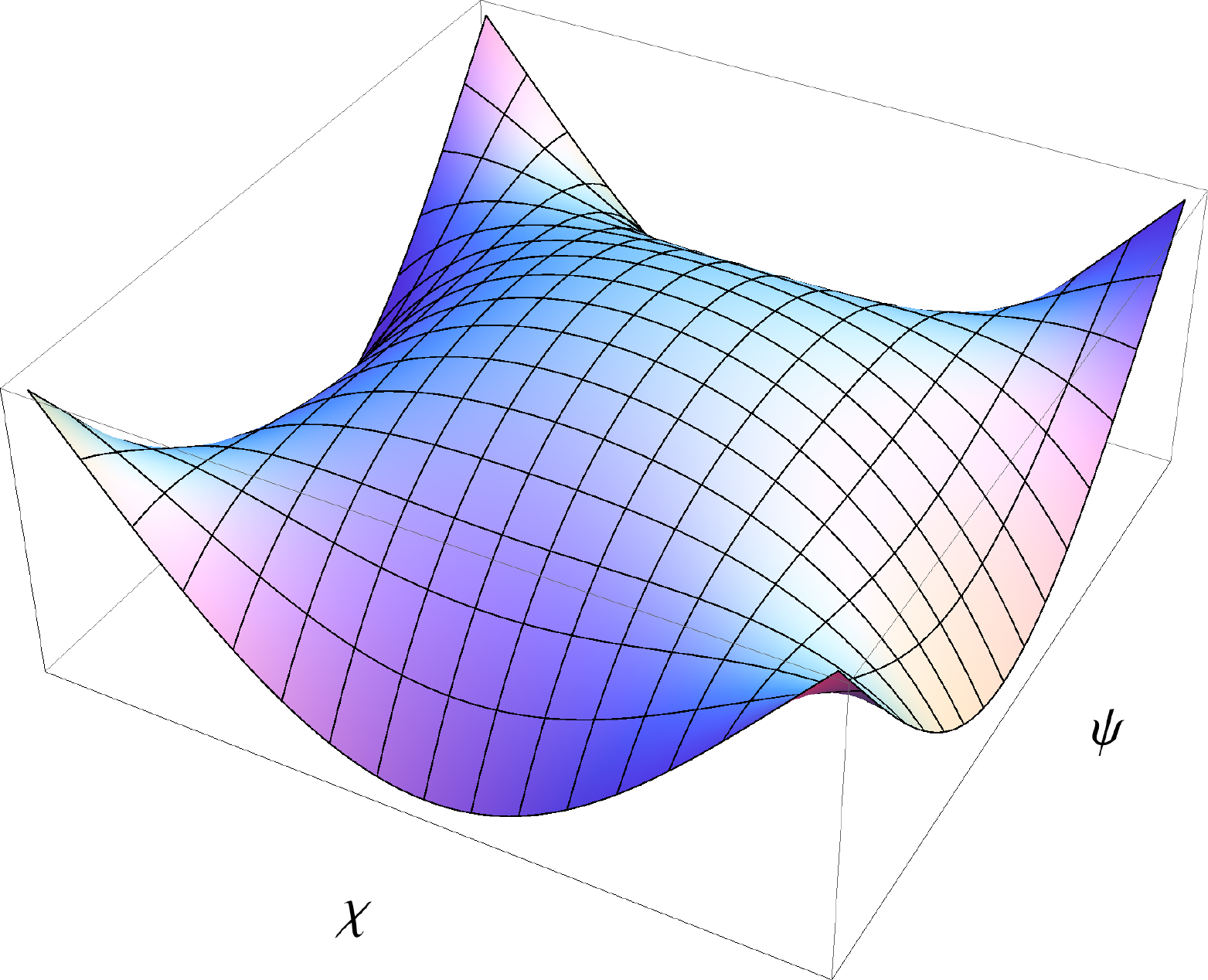}
\end{array}$
  \caption{Qualitative form of the potential \eqref{eq:Vscalar} for $\ell = 4$ on large scales (left) and zoomed in near the origin (right). It is easy to see that the potential is symmetric, and that we can restrict our discussion to the half-quadrant $\chi \geq \psi \geq 0$.}
  \label{fig:potential3D}
\end{figure}

Note that the inflaton potential \eqref{eq:Vscalar} is invariant under transformations $\Phi \rightarrow \Phi^*$ and $\Phi \rightarrow e^{2i\pi/\ell}\Phi$. These symmetries are also obvious in fig.~\ref{fig:potential3D} where the potential is plotted for $\ell = 4$. We can therefore assume that $0 \leq \psi \leq \chi \tan(\pi/\ell)$ without loss of generality, and we will do so throughout this paper.\footnote{For any $\chi$ and $\psi$, we can do a symmetry transformation $\Phi \rightarrow e^{2i\pi(n/\ell)}\Phi$ to get $\chi > 0$ and $-\chi \tan(\pi/\ell) < \psi \leq \chi \tan(\pi/\ell)$. If $\psi \geq 0$, our assumption is now valid. Otherwise, we can follow up with a symmetry transformation $\Phi \rightarrow \Phi^*$ to get $\psi > 0$.}

We only consider the case $\beta \geq 0$ in which the inflaton slow-rolls down the potential; for $\beta < 0$ the inflaton would be stuck at the local minimum at $\Phi=0$ and would have to tunnel to reach the global minimum.

\subsection{Construction from superpotential and \kahler potential}

Inflaton potentials of the form of eq.~\eqref{eq:Vscalar} can be constructed from different superpotentials and \kahler potentials. One possibility, used e.g.\ in \cite{flavonInflation,newInflationSUGRA2,HeisenbergSymmetry}, is the superpotential
\begin{align}
 W \, = \, \sqrt{V_0} \, S \left( 1 - 2^{\ell/2} \frac{ \Phi^\ell }{ M^2 } \right) \label{eq:eqW}
\end{align}
with the superfields $S$ and $\Phi$ and model parameters $V_0$, $M$ and $\ell$. $V_0$ and $M$ are chosen to be real, as their phases can be absorbed in the fields $S$ and $\Phi$. We use natural units with $8\pi G = M_{\text{Pl}}^{-2} = 1$ to keep the notation simple.

We also include the leading terms in an expansion of the \kahler potential
\begin{align}
 K \, = \, \left| S \right|^2 + \left| \Phi \right|^2 + \left( 1+\beta \right) \left| \Phi S \right|^2 - \kappa_S \left| S \right|^4 + ... \label{eq:eqK}
\end{align}
The scalar F-term potential for chiral superfields can be calculated from $W$ and $K$ with the formula
\begin{equation}
\label{eq:localSusyVF}
 V_F = e^K\left( D_i K^{i \overline{j}} D_j^*  - 3\lvert W \rvert^2  \right),
\end{equation}
where $K^{i \overline{j}}$ is the matrix inverse of the \kahler metric $K_{ \overline{i}j }$, and
\begin{align}
 K_{ \overline{i}j } = \dmd{K}{2}{ {X^*_i} }{}{ {X_j} }{},\quad
 D_i = \dpd{W}{ {X_i} } + W \dpd{K}{ {X_i} },\quad
 X = (S,\Phi).
\end{align}
For $W$ and $K$ given by eqs.~\eqref{eq:eqW} and \eqref{eq:eqK}, the scalar potential is\footnote{We do not include a D-term contribution, as we assume that inflation happens along a D-flat direction.}
\begin{align}
 V \, &\simeq \, V_0 \Bigl| 1 - 2^{\ell/2} \frac{ \Phi^\ell }{ M^2 } \Bigr|^2  + \frac{\ell^2 2^\ell V_0}{M^4} \left| S \Phi^{\ell-1} \right|^2  + V_0 ( 4\kappa_S \left| S \right|^2 - \beta \left| \Phi \right|^2  ). \label{eq:Vtempstep}
\end{align}
We assume that $\kappa_S > \frac{1}{12}$ so that the scalar component of $S$ has a mass above the Hubble scale $\Hubble$ and is stabilized at 0 during inflation. We can then neglect this field from now on. For the scalar component of $\Phi$, we choose the decomposition into real and imaginary part:
\begin{align}
 \Phi \, = \, \frac{1}{\sqrt{2}} \left( \chi + i \psi \right).
\end{align}
Inserting this decomposition in eq.~\eqref{eq:Vtempstep}, we arrive at the scalar potential given in eq.~\eqref{eq:Vscalar}.

Note that we only use the inflaton potential \eqref{eq:Vscalar} for the calculations, so our results are also valid for other superpotentials and \kahler potentials which lead to an effective inflaton potential of the form \eqref{eq:Vscalar}, e.g.\ for \cite{newInflationSUGRA1,preinflation}, and for \cite{newInflationSUGRA3} (in the limit $c \rightarrow 0$ in the last cited paper).

\section{Initial conditions}
In a multi-field model of inflation, the predictions can depend on the initial conditions. In this section, we want to discuss the initial conditions that we use for our analysis.

\subsection{$\Phi_{\text{i}} = 0$ from preinflation}

We start from the assumption that at some point in time, the field is very close to $\Phi = 0$. Such a state can be generated dynamically e.g.\ by a period of preinflation \cite{preinflation,newInflationSUGRA2}. Preinflation is a preceding phase of inflation, driven by some other inflaton field $\Xi$, during which $\Phi$ has a large mass from the vacuum expectation value of the inflaton field $\Xi$. This large mass drives $\Phi \rightarrow 0$ and keeps it stabilized there. After preinflation has ended, $\Xi \rightarrow 0$, and the $\Xi$-induced mass for $\Phi$ disappears. The inflaton $\Phi$ now sits at its ``initial value'' $\Phi=0$ and can start slow-rolling down its potential.

\subsection{Quantum diffusion boundary}

When $\Phi = 0$, the Friedmann equations imply that the field does not move away from $\Phi=0$ at all. However, the field $\Phi$ has quantum fluctuations which can be thought of as moving $\Phi$ randomly over time. Near $\Phi=0$, these quantum fluctuations dominate over the classical evolution; we call this area the quantum diffusion region. After some time, however, $\Phi$ has randomly walked away into some region where the potential is steep enough for the classical evolution to take over and the inflaton rolls away from $\Phi = 0$, down the potential gradient towards the nearest minimum. We therefore choose initial conditions on the boundary where the classical evolution starts to dominate over the quantum diffusion, and calculate the evolution with the Friedmann equations from there.

The boundary between the diffusion region and the region of classical field evolution can be estimated by comparing the classical evolution $\Delta \Phi_\text{cl}$ per Hubble time $t_H = \Hubble^{-1}$ to the growth of quantum fluctuations $\Delta \Phi_{\text{qu}}$ per Hubble time.
\begin{align}
 \lvert \Delta \Phi_\text{cl} \rvert^2 \, = \, \lvert  \dot{\Phi}  \rvert^2 \, t_H^2 \, = \, \frac{1}{V^2} \left[ \left( \frac{\partial V}{\partial \chi} \right)^2 + \left( \frac{\partial V}{\partial \psi} \right)^2 \right] \, \stackrel{!}{=} \, \lvert \Delta \Phi_\text{qu} \rvert^2 \, = \, 2\left( \frac{\Hubble}{2\pi} \right)^2, \label{eq:diffBoundaryIntroduce}
\end{align}
where we used the Friedmann equations
\begin{subequations}
\begin{align}
 \ddot{\chi} + 3 \Hubble \dot{\chi} + \partial_\chi V \, &= 0, \label{eq:Friedmann1}\\
 \ddot{\psi} + 3 \Hubble \dot{\psi} + \partial_\psi V \, &= 0, \label{eq:Friedmann2}
\end{align}
\end{subequations}
and
\begin{align}
 3\Hubble^2 \, &= \, V + \frac{\dot{\chi}^2}{2} + \frac{\dot{\psi}^2}{2}, \label{eq:Friedmann3}
\end{align}
in the slow-roll approximation $\ddot{\chi} \simeq \ddot{\psi} \simeq 0$, $3\Hubble^2 \simeq V$.

To solve eq.~\eqref{eq:diffBoundaryIntroduce}, it is more useful to work in polar coordinates for $\Phi$, with a radius $\phi$ and an angle $\theta$:
\begin{align}
 \Phi \, = \, \frac{1}{\sqrt{2}}( \chi + i \psi )  \,= \, \frac{\phi}{\sqrt{2}} \, e^{i \theta}.  \label{eq:polarCoord}
\end{align}
With these conventions, the scalar potential \eqref{eq:Vscalar} is
\begin{align}
 V \, = \, V_0 \left\{  1 - \frac{2}{M^2} \phi^\ell \cos( \ell \theta )  + \frac{\phi^{2\ell}}{M^4} - \frac{ \beta }{ 2 } \phi^2   \right\}. \label{eq:VscalarPolar}
\end{align}
We must also be careful to note that
\begin{align}
 \left( \frac{ \partial V }{ \partial \chi } \right)^2 + \left( \frac{ \partial V }{ \partial \psi } \right)^2 \, = \, \left( \frac{ \partial V }{ \partial \phi } \right)^2 + \left( \frac{1}{\phi} \frac{ \partial V }{ \partial \theta } \right)^2,
\end{align}
where the factor $1/\phi$ appears in front of the derivative with respect to $\theta$ because $\theta$ is not a canonically normalized field. In these polar coordinates, and using the fact that $V \simeq V_0$ during new inflation, eq.~\eqref{eq:diffBoundaryIntroduce} can be written as
\begin{align}
  \left( \frac{1}{V_0} \frac{ \partial V }{ \partial \phi } \right)^2 + \left( \frac{1}{V_0}\frac{1}{\phi} \frac{ \partial V }{ \partial \theta } \right)^2 \, = \, \left( \frac{\Hubble}{\sqrt{2}\pi} \right)^2. \label{eq:diffBoundary}
\end{align}
We want to solve this equation for the initial field values $\chi_i$ and $\psi_i$ separately for the cases $\beta = 0$ and $\beta \neq 0$.

\subsubsection{Diffusion boundary for $\beta = 0$}
For $\beta = 0$, the relevant derivatives are\footnote{We drop the term $\phi^{2\ell}/M^4$ in the potential because it is negligible during inflation.}
\begin{subequations}
\begin{align}
 \frac{1}{V_0}\frac{\partial V}{\partial \phi} \, &= \, -\frac{2 \ell}{M^2}  \phi^{\ell-1} \cos( \ell \theta ), \\
 \frac{1}{V_0}\frac{\partial V}{\partial \theta} \, &= \, \frac{2 \ell}{M^2}  \phi^{\ell} \sin( \ell \theta ).
\end{align}
\end{subequations}
With these ingredients, eq.~\eqref{eq:diffBoundary} becomes very simple
\begin{align}
 \left( \frac{2\ell}{M^2} \phi_i^{\ell-1} \right)^2 \, &= \, \left( \frac{\Hubble}{\sqrt{2}\pi} \right)^2,
\end{align}
so the diffusion boundary is a circle in field space with the squared radius
\begin{align}
 \left( \chi_i^2 + \psi_i^2 \right) \, &= \, \left( \frac{M^2 \Hubble}{\sqrt{8}\pi\ell} \right)^{ \frac{2}{\ell-1}}. \label{eq:diffBoundaryBeta0}
\end{align}

\subsubsection{Diffusion boundary for $\beta > 0$}
For sufficiently small $\Phi$, the mass term from $\beta$ dominates over the other inflaton potential terms which are of higher order in the fields. Therefore, we can neglect the other interactions at the diffusion boundary\footnote{If $\beta$ is so small that it does not dominate even at the diffusion boundary, it can be neglected completely for the purpose of inflation, and the results for $\beta = 0$ can be used.}, and eq.~\eqref{eq:diffBoundary} simplifies to
\begin{align}
 \left( \chi_i^2 + \psi_i^2 \right) \, = \, \left( \frac{\Hubble}{\sqrt{2}\pi\beta} \right)^2. \label{eq:diffBoundaryBeta}
\end{align}
We find that the diffusion boundary is a circle around the origin with radius $\Hubble/(\sqrt{2}\pi\beta)$.

To find out how large $\beta$ must be so that we can neglect the other inflaton interactions, we can compare the size of the two interactions at the diffusion boundary given by eq.~\eqref{eq:diffBoundaryBeta}:
\begin{align}
  &\frac{ \left[ \left( \frac{ \partial V }{ \partial \phi } \right)^2 + \left( \frac{1}{\phi} \frac{ \partial V }{ \partial \theta } \right)^2 \right]_{\beta = 0} }{ \left[ \left( \frac{ \partial V }{ \partial \phi } \right)^2 + \left( \frac{1}{\phi} \frac{ \partial V }{ \partial \theta } \right)^2 \right]_{\beta > 0} } 
  \, = \, \frac{ \left( \frac{2\ell}{M^2} \phi_i^{\ell-1} \right)^2 }{ \beta^2 \phi_i^2 } 
  \, = \, \frac{4\ell^2}{\beta^2M^4} \left( \frac{\Hubble}{\sqrt{2} \pi \beta} \right)^{2\ell-4}
  \, \ll \, 1 \notag \\
  \Leftrightarrow ~~~ &\beta \, \gg \, \left[  \frac{4\ell^2}{M^4} \left(  \frac{V_0}{6\pi^2}  \right)^{\ell-2}  \right]^{\frac{1}{2\ell-2}}.\label{eq:minBetaForBoundary}
\end{align}
For an order-of-magnitude estimate, we can insert the vacuum energy for single-field new inflation with $\beta \simeq 0$, which is
\begin{align}
 V_0 \, \sim \, 12\pi^2 A_s^2 \left( \frac{M^2}{2\ell} \right)^{ \frac{2}{\ell-2} } \left[  (\ell-2)N_e  \right]^{ -\left( 2 + \frac{2}{\ell-2} \right)}.
\end{align}
Then the condition \eqref{eq:minBetaForBoundary} becomes
\begin{align}
 \beta \, \gg \, \frac{ \left(  \sqrt{2} A_s  \right)^{ \frac{\ell-2}{\ell-1} } }{(\ell-2)N_e},
\end{align}
and we find that eq.~\eqref{eq:diffBoundaryBeta} is valid for any $\beta \gg 10^{-5}$ if $\ell=4$. For larger $\ell$, the threshold for $\beta$ is even lower.

\section{Analytic estimate of field trajectory}
In this section, we want to discuss the dynamics of the $\chi$ and $\psi$ fields during inflation to estimate for which parameters the model reduces to the well-known single-field model of new inflation.

We can rewrite the fields $\chi$, $\psi$ as $\phi$, $\theta$ according to eq.~\eqref{eq:polarCoord}. Using the partial derivatives
\begin{align}
 \frac{\partial \phi}{\partial \chi} = \cos(\theta),
 \quad\quad \frac{\partial \phi}{\partial \psi} = \sin(\theta),
 \quad\quad \frac{\partial \theta}{\partial \chi} = -\frac{ \sin(\theta) }{\phi},
 \quad\quad \frac{\partial \theta}{\partial \psi} = \frac{ \cos(\theta) }{\phi},
\end{align}
and ignoring the negligible term proportional to $\phi^{2\ell}$ in the potential, we can calculate the partial derivatives of the scalar potential \eqref{eq:VscalarPolar} during inflation:
\begin{subequations}
\begin{align}
 \partial_\chi V \, &= \, -V_0 \, \left\{ \beta \chi  +  \frac{2}{M^2} \frac{\partial}{\partial \chi}( \phi^\ell \cos(\ell\theta) )   \right\}   \notag\\
 &= \, -V_0 \, \chi \left\{  \beta  +  \frac{2\ell \phi^{\ell-2}}{M^2} \left(  \cos(\ell\theta)  +  \sin(\ell\theta)\tan(\theta)   \right) \right\},
 \label{eq:partialVchi} \\
 \partial_\psi V \, &= \, -V_0 \, \left\{ \beta \psi  +  \frac{2}{M^2} \frac{\partial}{\partial \psi}( \phi^\ell \cos(\ell\theta) )   \right\}  \notag\\
 &= \, -V_0 \, \psi \left\{  \beta  +  \frac{2\ell \phi^{\ell-2}}{M^2} \left(  \cos(\ell\theta)  -  \sin(\ell\theta)\cot(\theta)   \right) \right\}. \label{eq:partialVpsi}
\end{align}
\end{subequations}

Using the Friedmann eqs.~\eqref{eq:Friedmann1}--\eqref{eq:Friedmann2} and the slow-roll approximation $\ddot{\chi} \simeq \ddot{\psi} \simeq 0$, we can calculate the inflaton trajectory in field space:
\begin{align}
 \frac{\partial \psi}{\partial \chi} \, = \, \frac{\dot{\psi}}{\dot{\chi}} 
 \, = \, \frac{-\partial_\psi V}{-\partial_\chi V} 
 \, = \, \left( \frac{\psi}{\chi} \right) \frac{ \beta M^2  +  2\ell \phi^{\ell-2} \left(  \cos(\ell\theta)  -  \sin(\ell\theta)\cot(\theta)   \right) }{ \beta M^2  +  2\ell \phi^{\ell-2} \left(  \cos(\ell\theta)  +  \sin(\ell\theta)\tan(\theta)   \right) }. \label{eq:eomPsiChi}
\end{align}
We will discuss the behaviour of the field trajectory for two distinct cases: for a vanishing mass term ($\beta = 0$), for which we will recover the single-field new inflation limit, and for a large mass term ($\beta \gtrsim 10^{-2}$), for which we show that the imaginary inflaton component cannot generally be neglected.

\subsection{Supergravity mass term vanishes ($\beta = 0$): new inflation limit}
If $\beta=0$, the tachyonic mass term for $\Phi$ exactly vanishes.\footnote{This can naturally happen if such a mass term is forbidden by a symmetry of the \kahler potential, e.g.\ a Heisenberg symmetry \cite{HeisenbergSymmetry}.} In this case, we can show that the imaginary component $\psi$ decays before the observable primordial fluctuations leave the horizon, and inflation reduces to single-field new inflation.

With $\beta=0$, eq.~\eqref{eq:eomPsiChi} simplifies to
\begin{align}
 \frac{\partial (\log \psi)}{\partial (\log \chi)} \, =  \frac{  \cos(\ell\theta) - \sin(\ell\theta)\cot(\theta)   }{  \cos(\ell\theta)  +  \sin(\ell\theta)\tan(\theta)  } .
\end{align}
As explained in section \ref{sec:scalarPotential}, we can assume that $0 \leq \theta \leq \pi/\ell$. Moreover, for large $\theta > \theta_\text{thr}$, we have $\partial_\chi V > 0$, so for such large $\theta$, $\chi$ is rolling back towards $0$ until the inflaton is so close to $\chi = \psi = 0$ that quantum diffusion dominates. Eventually, the inflaton will randomly diffuse out of this diffusion region. Every time it leaves the diffusion region with $\theta > \theta_\text{thr}$, it is pushed back, until it eventually leaves the diffusion region with $\theta < \theta_\text{thr}$.

Assuming $0 < \theta < \theta_\text{thr}$, we find that
\begin{align}
 \frac{\partial (\log \psi)}{\partial (\log \chi)} \, = \, \frac{  \cos(\ell\theta) - \sin(\ell\theta)\cot(\theta)   }{  \cos(\ell\theta)  +  \sin(\ell\theta)\tan(\theta)  } \, = \, \frac{ \tan( (1-\ell)\theta ) }{ \tan \theta } \, < \, -(\ell-1).
\end{align}
This implies that $\psi$ drops off faster than $1/\chi^{\ell-1}$:
\begin{align}
 \psi(\chi) \, < \, \psi_0 \left( \frac{\chi_0}{\chi} \right)^{\ell-1} \label{eq:psiChiDropoff}
\end{align}
for any initial values $\psi_0$ and $\chi_0$. If the initial displacement from $\psi=\chi=0$ is due to quantum fluctuations as in eq.~\eqref{eq:diffBoundaryBeta0}, then eq.~\eqref{eq:psiChiDropoff} implies that $\psi$ is always negligible at horizon crossing. One can estimate this from the single-field results for new inflation, where the normalization of the CMB amplitude ensures that between the diffusion boundary and horizon crossing, $\chi$ grows by a factor of $\left( \chi_* / \chi_i \right)^{\ell-1} = A_s^{-1}$. Then eq.~\eqref{eq:psiChiDropoff} implies that $\frac{\psi_*}{\psi_i} < A_s$, and so $\frac{\psi_*}{\chi_*} < A_s^{1+\frac{1}{\ell-1}} \sim 10^{-5}$, so at horizon crossing $\psi$ is already negligible compared to $\chi$.

We conclude that for $\beta=0$, the model reduces to single-field new inflation in $\chi$.

\subsection{Supergravity mass term relevant ($\beta \gtrsim 10^{-3}$)}
If $\beta > 0$, $\psi$ grows during the early stages of inflation. For sufficiently large $\beta$, $\psi$ can be significantly large when the primordial fluctuations leave the horizon, and therefore the dynamics of $\psi$ is expected to have an effect on the CMB spectrum.

\begin{figure}[tbp]
  \centering
\includegraphics[width=0.55\textwidth]{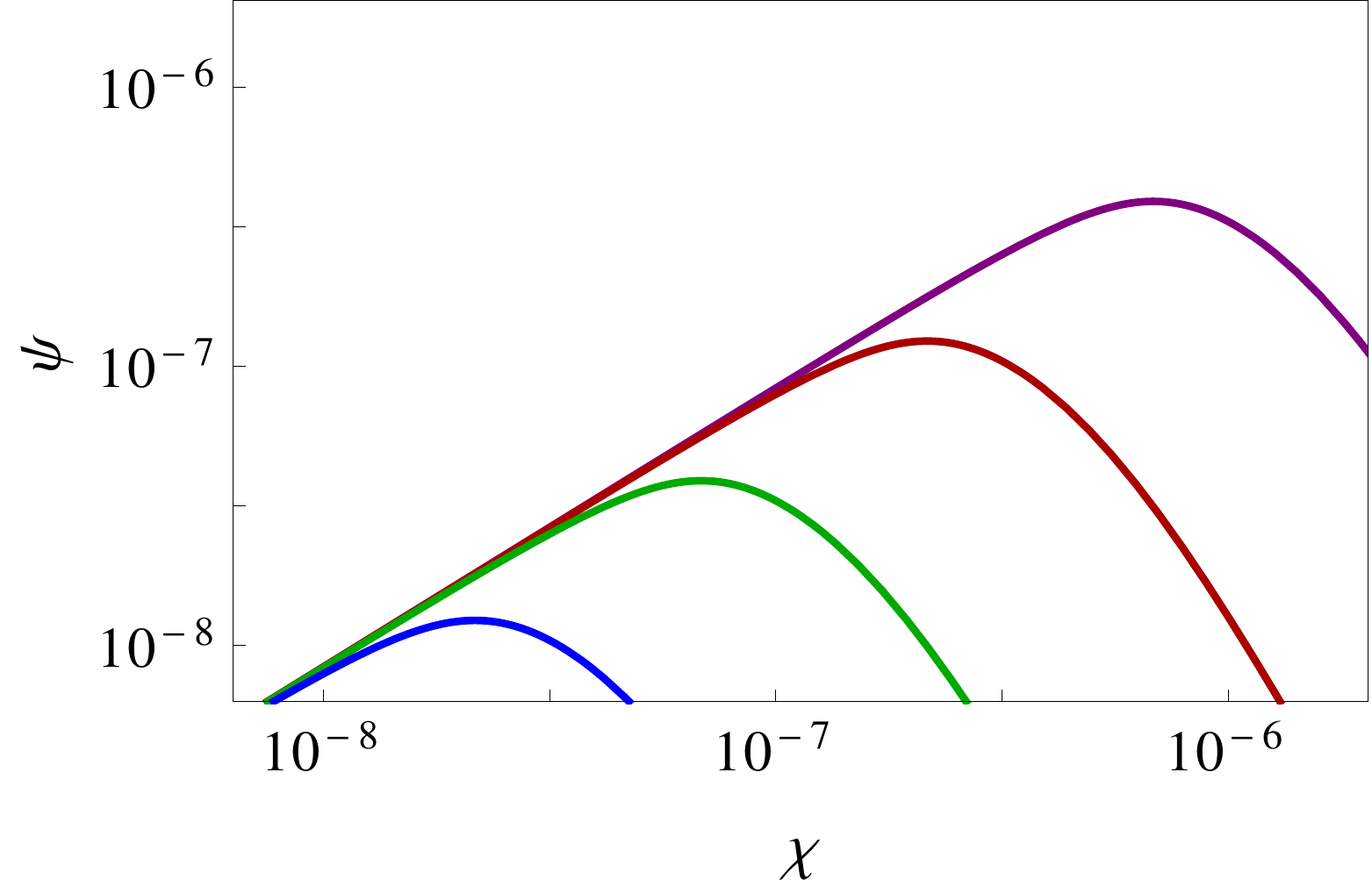}
  \caption{Field trajectories $\psi(\chi)$ for $\beta = 10^{-4}$ (blue), $\beta = 10^{-3}$ (green), $\beta = 10^{-2}$ (red) and $\beta = 10^{-1}$ (purple). The plots are shown for $\ell = 4$, $M = 10^{-5}$ and $\psi_i = (\tan 40^\circ) \chi_i$. We can clearly see both the growing phase, where $\beta$ dominates and $\psi$ grows linearly with $\chi$, and the decaying phase, where $\beta$ can be neglected and $\psi$ falls off as $\chi^{-3}$. Not surprisingly, the size of $\beta$ determines where the transition between the two phases takes place: for larger $\beta$, the growing phase lasts longer. Therefore, we can expect that deviations from single-field new inflation occur for large $\beta$, when the field $\psi$ does not decay before cosmological scales leave the horizon.}
  \label{fig:fieldTrajectory}
\end{figure}

For small field values $\phi^{\ell-2} \ll \beta M^2/(2\ell)$, $\psi$ grows proportionally with $\chi$, as we can see from eq.~\eqref{eq:eomPsiChi}:
\begin{align}
  &\frac{\partial \psi}{\partial \chi} \, \rightarrow \, \frac{\psi}{\chi} \quad \quad
  \Rightarrow \quad \psi \, \simeq \, \left( \frac{\psi_0}{\chi_0} \right) \chi. \quad \quad \text{(for $2\ell \phi^{\ell-2} \ll \beta M^2$)}
\end{align}
This linear growth of $\psi$ with $\chi$ continues until the fields grow large enough to make the higher-order interactions dominate over the tachyonic mass term from $\beta$. From that point onwards, the dynamics are approximately given by eq.~\eqref{eq:psiChiDropoff}: $\psi$ starts decaying faster than $\chi^{-(\ell-1)}$ while $\chi$ rolls towards the minimum at $\chi = M^{2/\ell}$. Both the growing and the decaying phase for $\psi$ can be clearly seen in fig.~\ref{fig:fieldTrajectory}.

If $\psi$ decays before cosmological scales leave the horizon, then we recover the single-field inflation limit with $\chi$ as the inflaton. We can use this to estimate for which values of $\beta$ we can expect significant deviations from the single-field limit. The transition from the growing to the decaying phase occurs at approximately $\chi_{\text{trans}}^{\ell-2} \sim \phi_{\text{trans}}^{\ell-2} \sim \beta M^2/(2\ell)$, as can be seen from eq.~\eqref{eq:eomPsiChi}. Calculating the field value $\chi_*$ at which cosmological scales leave the horizon is more involved. For a rough estimate, we assume that it is given by its single-field new inflation value:
\begin{align}
 \chi_*^{\ell-2} \, \simeq \, \frac{\beta M^2}{2\ell \left( \left(\frac{1 + (\ell-2)\beta}{1 - \beta}\right)  e^{(\ell-2)\beta N_e} - 1 \right)}.
\end{align}
where $N_e$ is the number of e-folds before the end of inflation at which cosmological scales leave the horizon. Setting $\chi_* \lesssim \chi_{\text{trans}}$, we find that multi-field effects are expected to become important for $\beta$ above the threshold value
\begin{align}
 \beta \, \gtrsim \, \frac{\ln(2)}{(\ell-2)N_e} \, \sim \, \frac{10^{-2}}{\ell-2}.
\end{align}
We therefore recover the single-field limit for $\beta \ll 10^{-2}/(\ell-2)$, whereas for $\beta \gtrsim 10^{-2}/(\ell-2)$ the predictions should be calculated in a multi-field formalism. This is done numerically in section~\ref{sec:numeric}.

\subsection{Generalization to initial field values outside the diffusion region}
\label{sec:initialConditionsOutsideDiffBoundary}

So far, we have assumed that inflation starts at $\Phi_i \simeq 0$ inside the diffusion region given by eq.~\eqref{eq:diffBoundaryBeta0} or \eqref{eq:diffBoundaryBeta}. Due to the field dynamics, our results can be immediately generalized to initial field values $\Phi_i$ outside the diffusion region, as long as $\Phi_i$ remains sufficiently small.

For $\beta>0$, we have $\psi \propto \chi$ and therefore $\theta = \text{const}$ for small field values. This implies that our results, which depend on $\theta_i$, are valid even when the initial value $\Phi_i$ is outside of the diffusion region, as long as $\Phi_i$ starts in the growing phase in which $\psi \propto \chi$. A sufficient condition for the validity of our results is $\phi_i^{\ell-2} \ll \frac{\beta M^2}{2\ell}$.

For $\beta=0$, no growing phase exists. However, because $\psi$ quickly decays according to eq.~\eqref{eq:psiChiDropoff}, we recover the single-field limit even for larger initial field values. We expect significant deviations from the single-field results only for initial values close to the horizon crossing scale, such that $\psi$ has very little time to decay between the start of inflation and horizon crossing.

\section{Numerical treatment}
\label{sec:numeric}
In the last section, we have discussed that the inflationary predictions can be changed by the imaginary inflaton component for $\beta \gtrsim 10^{-2}/(\ell-2)$. In this section, we present a numerical calculation of the primordial perturbations using the $\delta N$ formalism.

\subsection{$\delta N$ formalism}

A powerful tool for calculating the primordial perturbations in multi-field models of inflation is the $\delta N$ formalism \cite{deltaN:1,deltaN:2,deltaN:3,deltaN:4}. It is based on the fact that the curvature perturbation $\zeta(x,t)$ on a spatial uniform-density hypersurface is given by the difference $\delta N$ in the number of e-folds between a flat initial hypersurface and the uniform-density final hypersurface:
\begin{align}
 \zeta(t,x) \, = \, \delta N(t,x) \, = \, N(t,x) - N_0(t). \label{eq:zetaDeltaN}
\end{align}
We choose the initial flat hypersurface at the time $t_*$ at which cosmological scales leave the horizon. In this paper we want to calculate the perturbations at the end of inflation, so we choose the final hypersurface of uniform density at the end of slow-roll inflation.

As the field perturbations are very small, one can expand $\delta N$ in powers of the field perturbations $\delta \phi_i$ on the initial flat hypersurface (the index $i$ denotes the $i$-th inflaton field):
\begin{align}
 \zeta \, = \, \delta N \, = \, \sum_i N_i \, \delta \phi_i \, + \, \sum_{i,j} N_{ij} \, \delta \phi_i \, \delta \phi_j \, + \, O(\delta \phi^3), \label{eq:deltaNexpand}
\end{align}
where we introduced the notation
\begin{align}
 N_i \, = \, \frac{\partial N}{\partial (\delta \phi_i)}, \quad N_{ij} \, = \, \frac{\partial^2 N}{\partial (\delta \phi_i) \partial (\delta \phi_j)}.
\end{align}
Inserting the de Sitter space field perturbations in eq.~\eqref{eq:deltaNexpand}, one can derive the leading-order expression for the amplitude $A_s$ of the primordial curvature perturbation:
\begin{align}
 A_s^2 \, &= \, \frac{\Hubble_*^2}{4\pi^2} \sum_i N_i^2 \, = \, \frac{V_*}{12 \pi^2} \sum_i N_i^2, \label{eq:deltaNV0}
\end{align}
where a subscript star indicates that a quantity should be evaluated at the time of horizon crossing.

With some extra work, one can also calculate the amplitude $f_{\text{NL}}$ of the reduced bispectrum \cite{deltaN:3}
\begin{align}
 f_{\text{NL}} \, &= \, -\frac{5}{6} \frac{ \sum_{ij} N_i N_j N_{ij} }{ \left( \sum_i N_i^2 \right)^2 }, \label{eq:deltaNfNL}
\end{align}
and the spectral index $n_s$ of the curvature perturbation \cite{deltaN:2}
\begin{align}
 n_s \, 
 &= \, 1 - 2 \varepsilon_* - \frac{2 \sum_{ij} \left( \frac{\partial V}{\partial \phi_i} \right)_* N_j N_{ij} }{ V_* \sum_i N_i^2 }, \label{eq:deltaNns}
\end{align}
with the first slow-roll parameter
\begin{align}
\varepsilon \, &= \, -\frac{\dot{\Hubble}}{\Hubble^2} \, \simeq \, \frac{1}{2V^2} \sum_i \left( \frac{ \partial V }{ \partial \phi_i } \right)^2.
\end{align}

For calculating the predictions for any set of parameters and initial conditions $(\chi_i, \psi_i)$, we took the following steps:
\begin{enumerate}
 \item We integrated the slow-roll equations of motion starting from $(\chi_i, \psi_i)$ forward in time until slow-roll inflation ends at $\lvert \eta \rvert = 1$. From this trajectory, we determined the end-of-inflation energy density $\rho_{\text{end}}$ and the background field values $(\chi_*, \psi_*)$ at $N_e  = 55$ e-folds before the end of inflation.
 \item For very small displacements $\Delta \chi$ and $\Delta \psi$, we integrated the equations of motion without using the slow-roll approximation, starting from $(\chi_* \pm \Delta \chi, \psi_* \pm \Delta \psi)$ and ending on the final uniform-density hypersurface with $\rho = \rho_{\text{end}}$, to determine the number $N$ of e-folds along these trajectories.
 \item We calculated the first and second derivatives from the difference quotients, e.g.\ $N_\chi = \frac{N(\chi_* + \Delta \chi, \psi_*) - N(\chi_*, \psi_*)}{\Delta \chi}$.
 \item We used eqs.~\eqref{eq:deltaNV0}--\eqref{eq:deltaNns} to calculate the primordial spectrum from the $N_i$, $N_{ij}$, $\chi_*$ and $\psi_*$.
\end{enumerate}

\subsection{Numerical results}

\begin{figure}[tbhp]
  \centering$
\begin{array}{cc}
\includegraphics[width=0.48\textwidth]{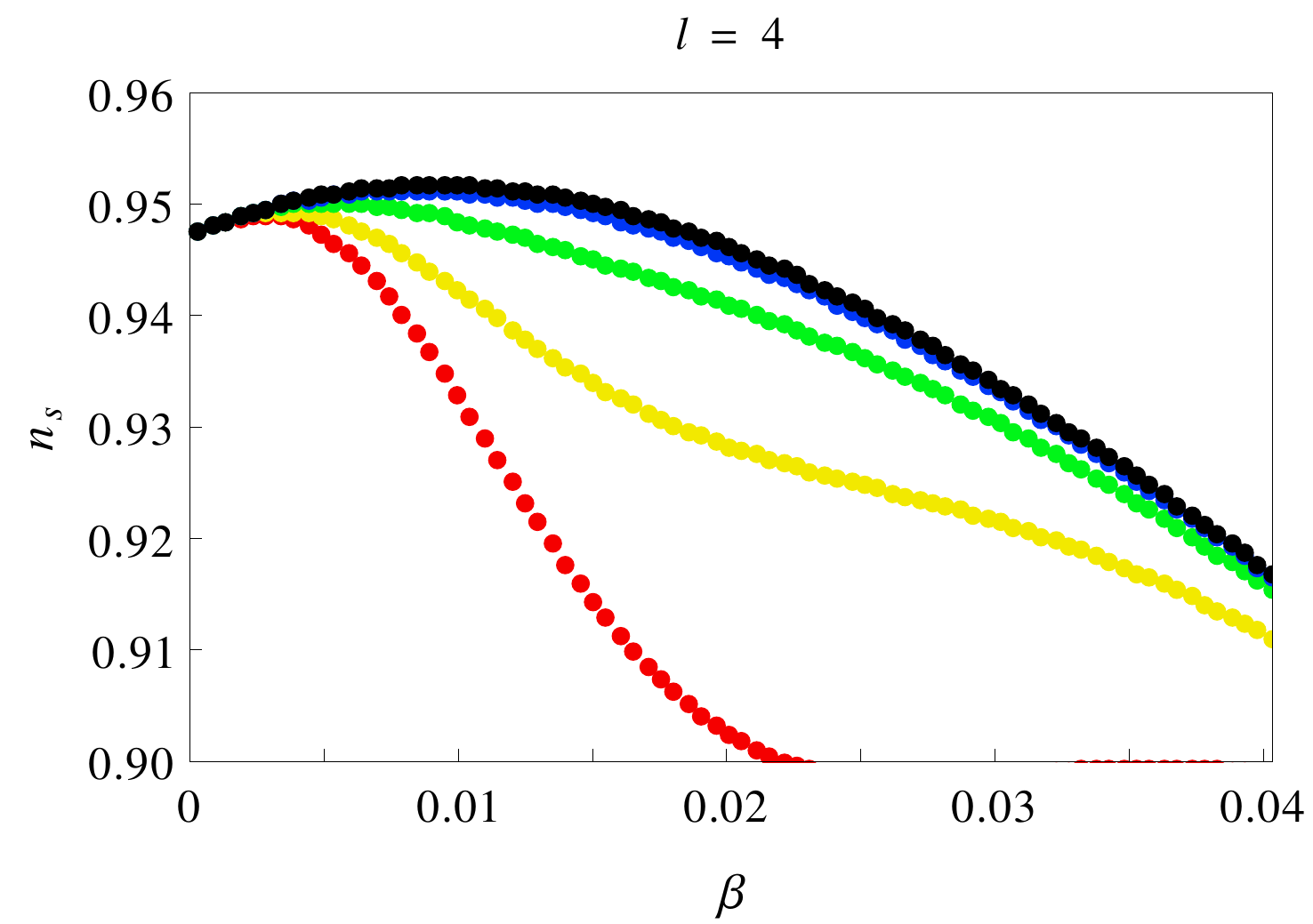} &
\includegraphics[width=0.48\textwidth]{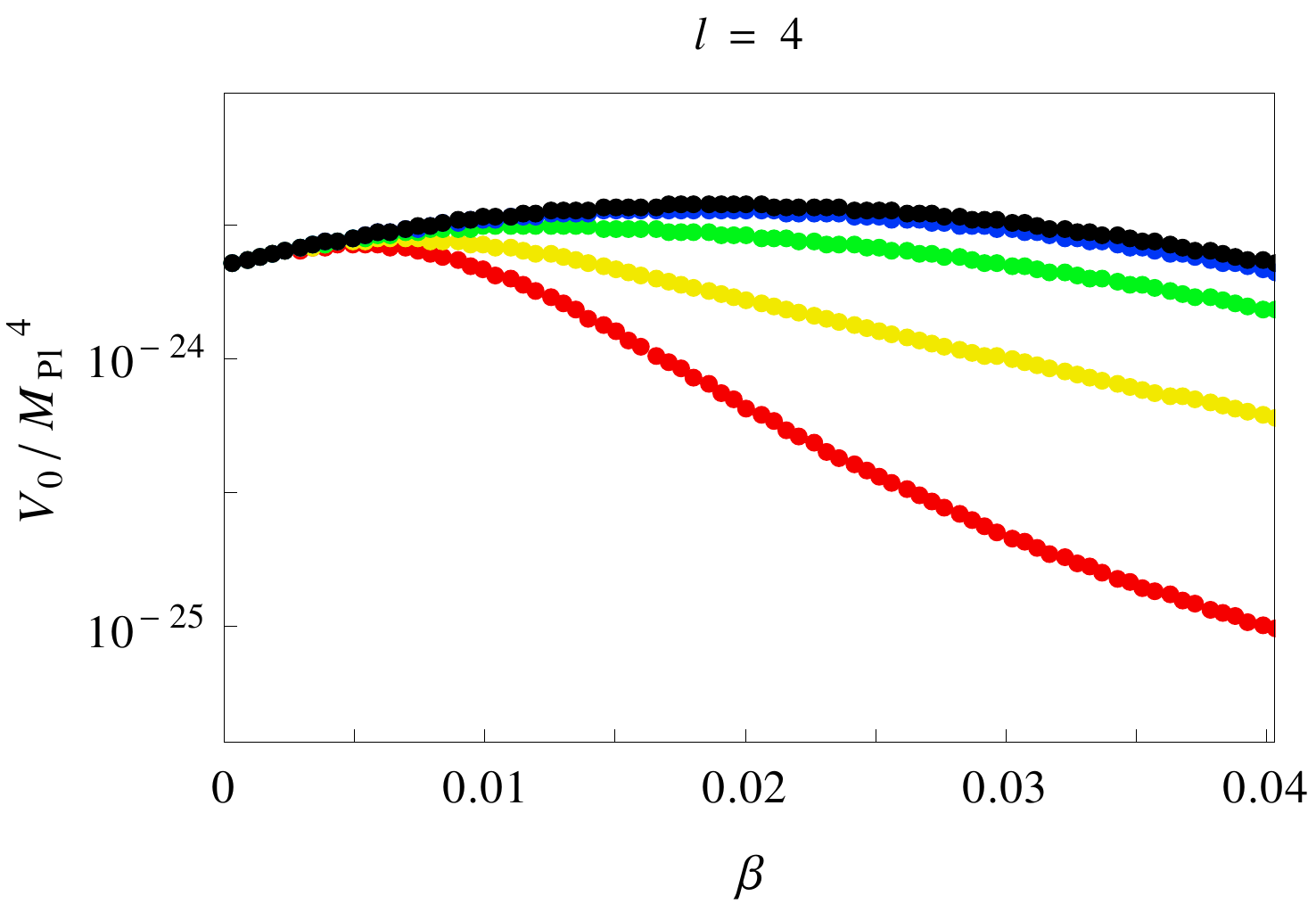} \\
\includegraphics[width=0.48\textwidth]{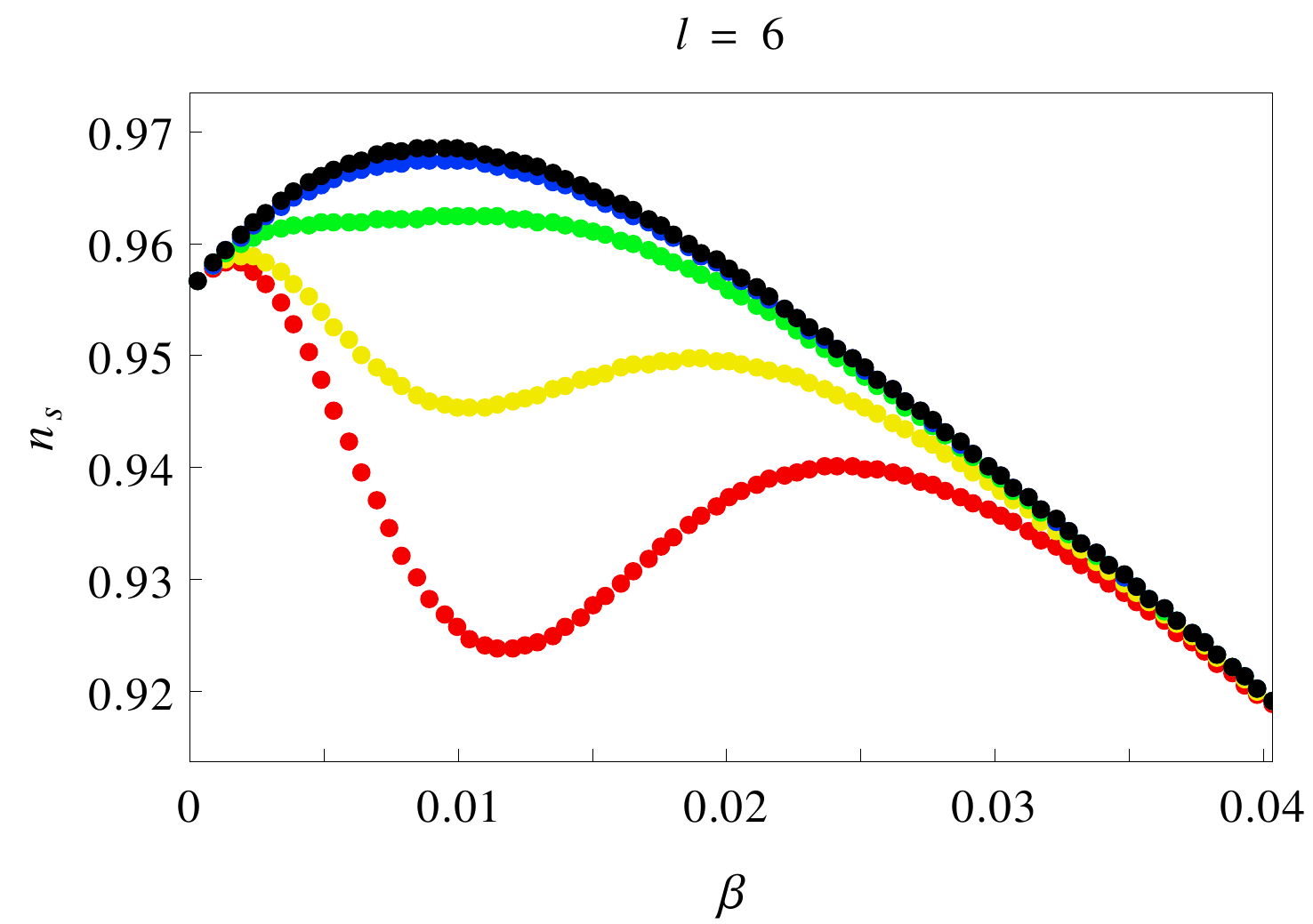} &
\includegraphics[width=0.48\textwidth]{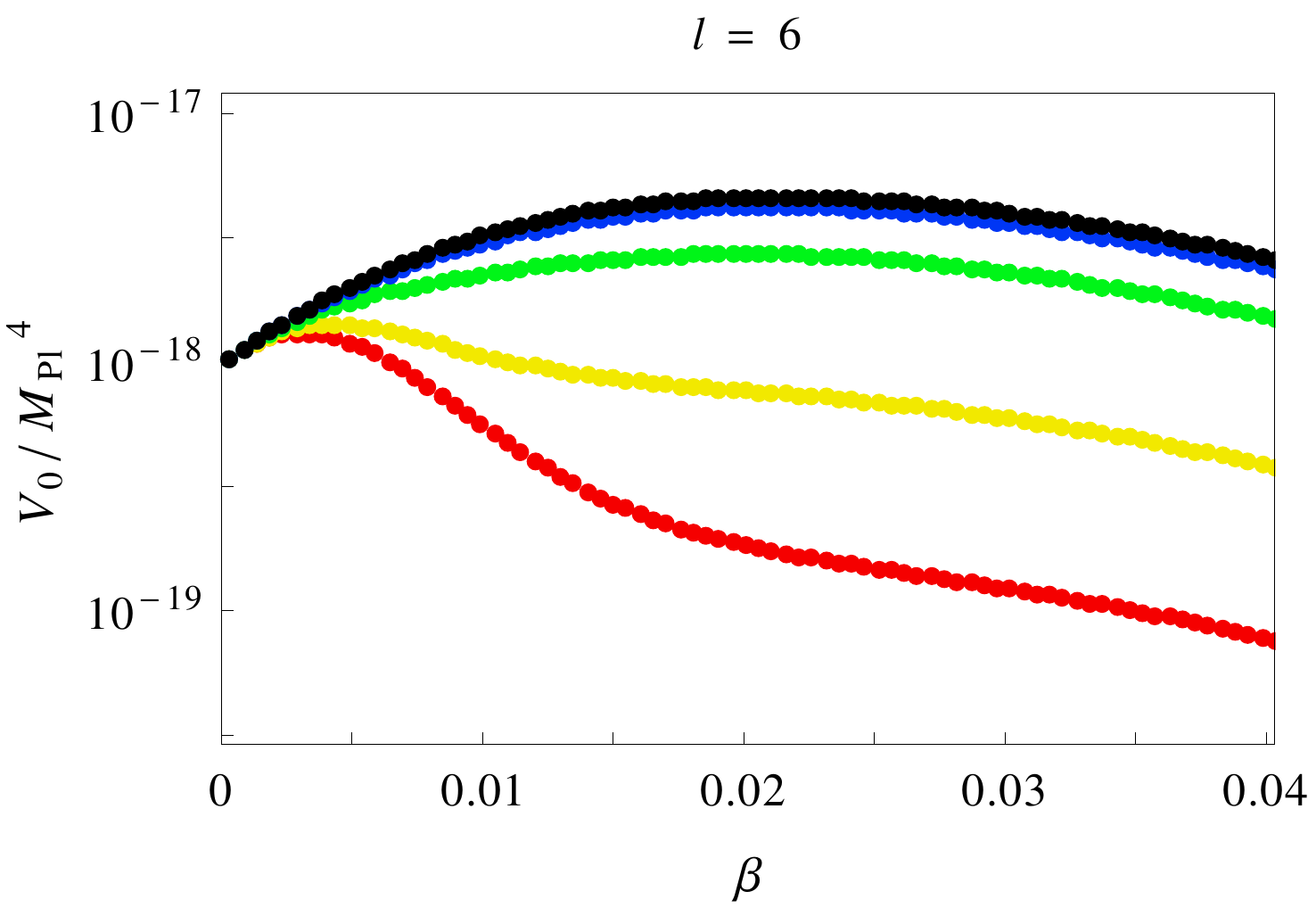} \\
\includegraphics[width=0.48\textwidth]{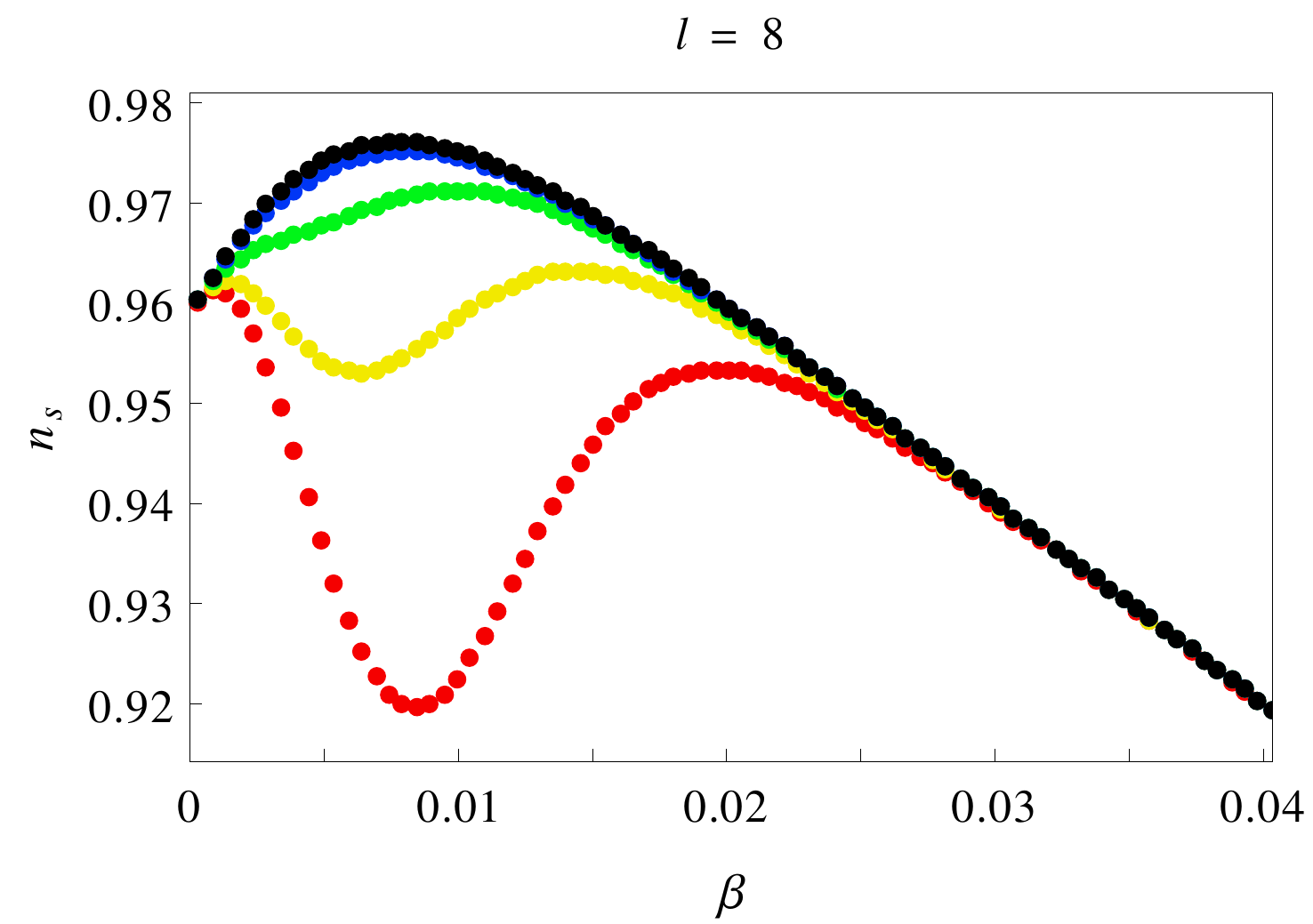} &
\includegraphics[width=0.48\textwidth]{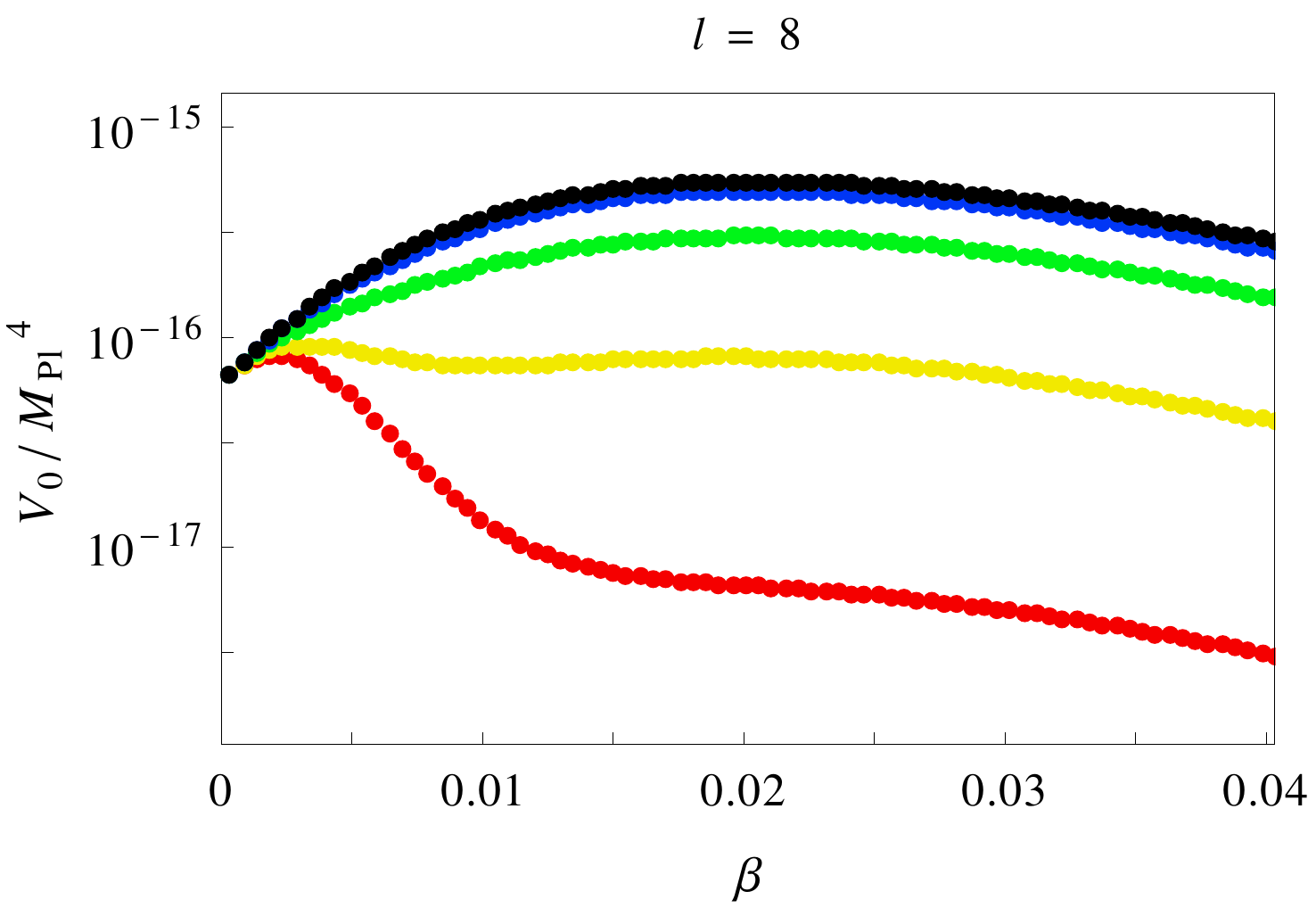}
\end{array}$
  \caption{Spectral index $n_s$ and vacuum energy $V_0$ as a function of $\beta$ for $\ell = 4$ (upper row), $\ell = 6$ (middle row) and $\ell = 8$ (bottom row). All plots are done for $N_e = 55$ and $M = 10^{-5}$; changing $M$ only introduces an overall factor in $V_0$ but does not change its $\beta$-dependence, while $n_s$ is completely independent of $M$. The differently coloured dots correspond to different initial angles $\theta$ in the $\chi$-$\psi$-plane: black is the single-field limit ($\theta = 0^\circ$) and red is close to the maximum angle ($\theta = 180^\circ/\ell$). The angles shown are $0^\circ$, $15^\circ$, $30^\circ$, $40^\circ$ and $44^\circ$ (for $\ell=4$), $0^\circ$, $10^\circ$, $20^\circ$, $27^\circ$ and $29^\circ$ (for $\ell=6$) and $0^\circ$, $7.5^\circ$, $15^\circ$, $20^\circ$ and $22^\circ$ (for $\ell = 8$). We see that increasing $\theta$ reduces $n_s$ and $V_0$, except for $\beta \simeq 0$ which always reproduces the single-field result. The effect is mild for most angles, and only close to the maximum angle the deviation becomes significant.}
  \label{fig:nsV0Numerics}
\end{figure}

\begin{figure}[tbp]
  \centering
$\begin{array}{cc}
\includegraphics[width=0.48\textwidth]{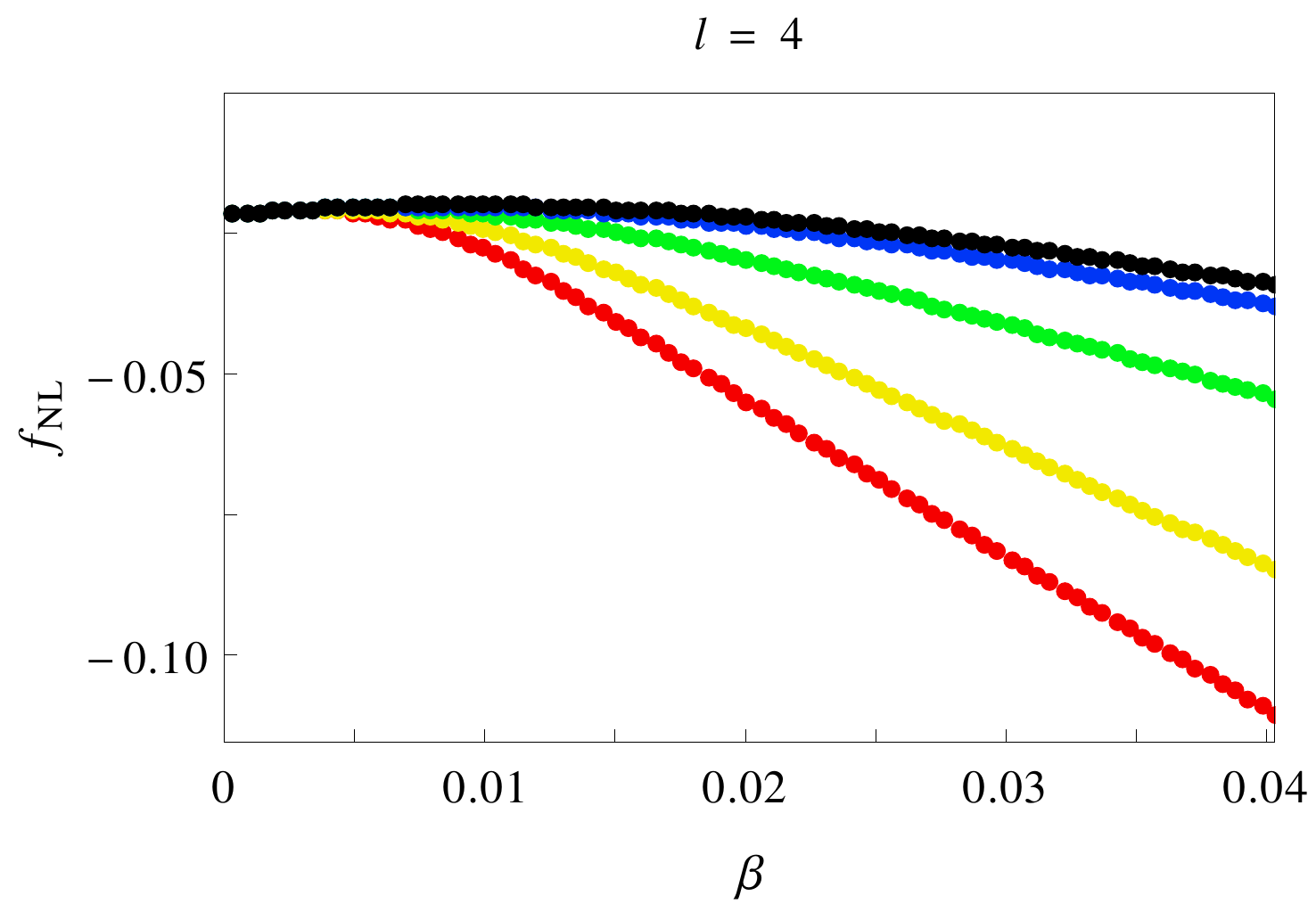} &
\includegraphics[width=0.48\textwidth]{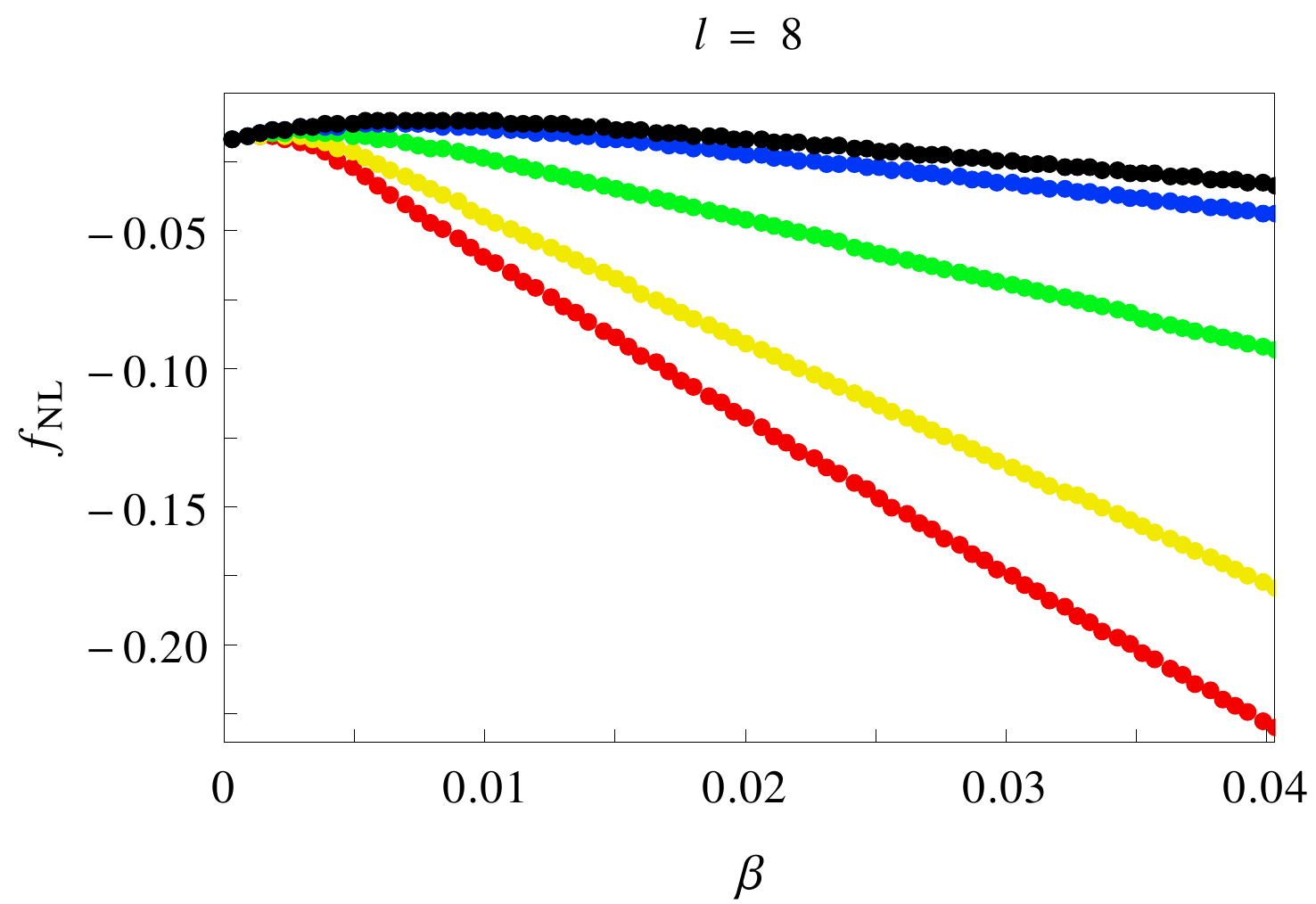}
\end{array}$
  \caption{Reduced bispectrum $f_{\text{NL}}$ as a function of $\beta$ for $N_e = 55$, $M = 10^{-5}$ and $\ell = 4$ (left) or $\ell = 8$ (right). The differently coloured dots correspond to different initial angles $\theta$ in the $\chi$-$\psi$-plane: black is the single-field limit ($\theta = 0^\circ$) and red is close to the maximum angle ($\theta = 180^\circ/\ell$). The angles shown are $0^\circ$, $15^\circ$, $30^\circ$, $40^\circ$ and $44^\circ$ (for $\ell=4$) and $0^\circ$, $7.5^\circ$, $15^\circ$, $20^\circ$ and $22^\circ$ (for $\ell = 8$). Independently of $M$, we generally find $\lvert f_{\text{NL}} \rvert < 1$, which makes it practically indistinguishable from zero.}
  \label{fig:fNL}
\end{figure}

Using the $\delta N$ formalism, we have calculated the spectral index $n_s$, the amplitude of the reduced bispectrum $f_{\text{NL}}$ and the vacuum energy $V_0$ during inflation. We assumed that inflation starts on the circle given by eq.~\eqref{eq:diffBoundaryBeta} where the classical evolution starts to dominate over the quantum diffusion of the inflaton fields.\footnote{Eq.~\eqref{eq:diffBoundaryBeta} is valid only for $\beta \gg 10^{-5}$, which is satisfied for all points in figs.~\ref{fig:nsV0Numerics} and \ref{fig:fNL}. For smaller $\beta$, one recovers the single-field limit, and no numerical calculation is needed.} We calculated the predictions for various points on this initial surface, parametrized by the angle
\begin{align}
  \theta \, = \, \arctan(\psi_i / \chi_i),
\end{align}
where the maximum angle is
\begin{align}
  \theta_{\text{max}} \, = \, \pi / \ell
\end{align}
because as we explained in section \ref{sec:scalarPotential}, larger angles are related to angles in the range $0 \leq \theta \leq \theta_{\text{max}}$ by symmetry transformations, so we can restrict ourselves to angles between 0 and $\theta_\text{max}$.

The results for $n_s$ and $V_0$ are shown in fig.~\ref{fig:nsV0Numerics} for $N_e = 55$ and $M=10^{-5}$. The numerical results confirm that we recover the single-field limit for $\beta \rightarrow 0$, while for larger $\beta$ the imaginary inflaton component reduces both $n_s$ and $V_0$. Note that the green dots, which are quite close to the black single-field result, correspond to $\theta = \frac{2}{3}\theta_{\text{max}}$. Therefore, even though the deviations become large for maximal $\theta$, most initial conditions give results similar to the single-field limit.

$f_{\text{NL}}$ is shown in fig.~\ref{fig:fNL}. It is generally in the range $-1 < f_{\text{NL}} < 0$ which is too small to be observed, even for close-to-maximal $\theta$.

We have checked that the $\beta$-dependence of our results is insensitive to changes in $M$. As in the single-field case, different choices of $M$ only give a constant factor for $V_0$, while $n_s$ does not depend on $M$ at all.

Note that while the results depend sensitively on $\theta_i$, they are valid for any initial $\phi_i = \sqrt{\chi_i^2 + \psi_i^2}$ as long as it is sufficiently close to zero (see section~\ref{sec:initialConditionsOutsideDiffBoundary}).

\section{Summary and conclusions}

In this paper, we have studied the effects of the multi-field dynamics of the complex scalar inflaton field in supergravity new inflation, providing a brief analytical discussion and a numerical calculation. We have found that for most of the parameter space, the model is well described by the usual single-field approximation, where only the real component of the inflaton is considered and its imaginary component is set to zero. In particular, this is the case if the mass term from the \kahler potential is very small or absent, in which case the imaginary component is quickly driven to zero before cosmological scales leave the horizon.

For a sufficiently large mass term, the results become sensitive to the initial conditions. For these cases, we have numerically calculated the predictions using the $\delta N$ formalism. For most initial conditions, we find that the results are still similar to the single-field results, but the deviations become significant for large initial values of the imaginary inflaton component (see fig.~\ref{fig:nsV0Numerics}). Those deviations generally reduce the spectral index $n_s$ and the inflationary vacuum energy $V_0$ compared to the single-field case. The reduced bispectrum is within the range $-1 < f_{NL} < 0$, which is in good agreement with the current data from the Planck experiment, but probably too small to ever be observed.

Our conclusions are twofold. First, we want to emphasize that new inflation in supergravity is well-approximated by a single-field model if either
\begin{itemize}
 \item the mass term of the inflaton is very small compared to the Hubble scale ($m_\Phi^2 \ll \Hubble/100$), or
 \item the initial displacement of the imaginary component is small (up to about 1/2 of its maximum value).
\end{itemize}
Second, if both of these conditions are violated (if both the inflaton mass term and the initial value of the imaginary inflaton component are sufficiently large), the spectral index $n_s$ and the vacuum energy $V_0$ depend sensitively on the initial conditions, and the single-field results should be interpreted as upper limits on $n_s$ and $V_0$ only.

\section*{Acknowledgements}
I want to thank Stefan Antusch and Stefano Orani for many valuable discussions.

\end{document}